\documentstyle[prl,aps,multicol]{revtex}

\begin{document}
\title{Dynamical stripe correlations and the spin fluctuations in cuprate
superconductors}

\author{J. Zaanen and W. van Saarloos\address{Institute Lorentz for
Theoretical Physics, Leiden University, \\ P.O. Box 9506, NL-2300 Leiden,
The Netherlands} }
\date{\today ; E-mail: jan@lorentz.leidenuniv.nl,
saarloos@lorentz.leidenuniv.nl}
\maketitle

\begin{abstract}
It is conjectured that the anomalous spin dynamics observed in the normal
state of cuprate superconductors might find its origin in a nearly ordered
spin system which is kept in motion by thermally meandering charged domain
walls. `Temperature sets the scale' finds a natural explanation, 
while a crossover to a low temperature quantum domain wall fluid is implied. 
\end{abstract}

\begin{multicols}{2}
\narrowtext
\section{Introduction}

Much excitement is generated by the observation by Tranquada {\it et al.}
of the striped phase in the cuprates\cite{tran}, which appears to be
in a tight competition with the superconducting state as long as
the doping level is not too high\cite{tran1}. Taken together with
the observation of the persistent gap in the normal state\cite{photoemission}
and the finding that a large magnetic field stabilizes an insulating
state\cite{boebinger}, this points to a bosonic physics. The electrons
pair up above $T_c$, forming an interacting boson system and as in, e.g.,
$^4He$ the crystalline (`charge-ordered') and superconducting state appear
as the natural candidates for the ground state. 

The real novelty of the striped phase is that it is at the same
time a N\'eel spin condensate. At  least, it is a bound state of
charge and spin\cite{zagu}, with the specialty that
the charge sector couples into the spin sector in the form of {\em disorder
operators}: the charged stripes are anti-phase boundaries
in the N\'eel state. In fact, all evidence for persisting dynamical stripe
correlations in the superconductors rests entirely on the spin
sector\cite{tran1}. 

Initially we considered the possibility of stripe-like correlations, 
motivated by the anomalous behavior of the
spin-fluctuations in the cuprates\cite{zhvs}. Both inelastic neutron
scattering\cite{aeppli,aeppli1} and NMR\cite{nmr} reveal that these
fluctuations show a characteristic time- ($\tau$)
and length-scale ($\xi$) 
which are both precisely inversely proportional to temperature.
The lack of an intrinsic scale is reminescent of a (quantum) critical regime.
This is the underlying idea 
behind  the `nearly antiferromagnetic Fermi liquid' idea
of Pines and coworkers\cite{pines}: the system is on its way
to a zero temperature ordered spin state, and superconductivity intervenes
at the last moment. 

There is a conceptual difficulty with this scenario. 
It is expected that temperature and doping play the role of
control parameters: how can it be that the critical regime extends over
a temperature- and doping range of $\sim 1000$ K and $\sim 10$ \%, 
respectively? At least
classical critical regimes do not behave in this way. 
Assuming that the system is on its way to the striped phase, the
relative insensitivity to doping dependence is trivially explained. In
contrast to normal antiferromagnets, the striped phase is an incommensurate
solid which can exist over a large doping range. The non-trivial part is
to explain the temperature dependences of the characteristic dynamical
scales.

The problem of thermally fluctuating domain walls 
in two dimensions (`incommensurate fluids') has been studied in great 
detail\cite{copper}. In such systems, the phenomenon of
{\em entropic  repulsions} is common. 
For increasing temperature, the thermal meandering
motions of individual domain walls increases, as well as the number of
collisions between domain walls. Collisions cost entropy and
the net effect is that the stiffness of the system as a whole increases. 
Due to these entropic repulsion effects, temperature becomes an
important parameter in the dynamics. It is emphasized that the basic effect 
is unrelated to criticality and it can therefore appear over a large 
temperature range. As we showed in an earlier paper\cite{zhvs}, the behaviors
of the characteristic time- and length scales of the spin fluctuations 
match in a natural way with the expectations for a thermally fluctuating
incommensurate fluid in two dimensions.  

Although the essence of our earlier work is unchanged, the relationship
between the domain wall dynamics and the observed spin dynamics was treated
too casually in this earlier paper\cite{zhvs}. After summarizing
the basics, we will make this connection more precise. This is used,
together with a refined experimental characterization of the normal
state magnetic fluctuations\cite{aeppli1},
to deduce a more complete picture of the domain wall fluid. 

\section{Incommensurate fluids in two dimensions.}

The standard theory of incommensurate fluids\cite{copper} starts
with the assumption that single domain walls (DW) behave like Gaussian
strings. It is assumed that its motions can be completely parametrized 
in terms of a transversal sound mode $\omega_q = c_W q$, where $q$ is the
wave number along the string, while $c_W = \sqrt{\Sigma / \rho}$ ($\Sigma$
is the string tension and $\rho$ the mass density). In addition, it is 
convenient to first only consider hard core string-string interactions.
At finite temperatures, DW will be subject to meandering motions. The
mean square fluctuation of a single string in 2D in the transverse $z$
direction between two points separated by a distance $l$ diverges as
$\langle ( \Delta z_l )^2 \rangle \simeq { {k_B T} \over {\Sigma} } l$.
When the walls have an average distance $d$, they will on average 
collide when $\langle (\Delta z_{l_c})^2 \rangle \simeq d^2$.
The typical distance between collision points follows immediately,
\begin{equation}
l_c \simeq { {\Sigma d^2} \over {k_B T} } 
\end{equation}  
This is the most important length scale in this problem. It sets (a) the
crossover length where the system changes from single wall dynamics to
that of the domain wall fluid, (b) it determines the collision density and
thereby the strength of the entropic repulsions, and (c) it corresponds
to the shortest spin-spin disordering length in the special case of the
striped fluid. For a further elaboration of the statistical physics we
refer to the literature\cite{copper}. 
The essence is that the domain
wall system can be described as a 2D elastic medium with an elastic constant
which itself is proportional to temperature: $F \sim T \int d\vec{r}
(\Delta \vec{n} )^2$. This is like a XY spin system with $J \sim T$, and
it follows that it is characterized by a {\em zero temperature
Kosterlitz-Thouless transition}: 
at every finite temperature, free dislocations proliferate and the system
is a fluid.

In order to adress the temporal aspects of the dynamics we suggested that
the single wall dynamics is {\em coherent} on length scales $\leq l_c$: the
single string sound oscillations are assumed to be underdamped. If this is the
case, the relative motion of two points a distance $l$
apart becomes important after a time $l/c_W$, and likewise the mean average
fluctuations of two points a distance $l_c$ apart deviates significantly
from the value d after a time of order $l_c/c_W$. Using Eq. (1), it follows
that the collision frequency is of order\cite{zhvs},
\begin{equation}
\hbar \Gamma = \mu k_B T, \;\; \mu = { {\pi \hbar} \over {\rho c_W d^2} }
\end{equation}
Hence, we find both a characteristic length- and time
scale which vary as $1/T$, with the obviously important meaning that
they both refer to the crossover from single wall to many wall behavior.
It remains to be demonstrated that these characteristic scales govern 
the spin dynamics as well.   
      
\section{Charged domain walls and spin fluctuations.}

The unique feature of the stripes is that it corresponds with an incommensurate
solid/fluid problem in the charge/longitudinal sector, coupled into a
spin problem: when the stripes are ordered one is still left  with
a Heisenberg antiferromagnet. Besides the modes of the previous
section, one has to deal with the transversal fluctuations associated
with the rotational invariance of the spin system. At least on the classical
level, {\em the domain wall- and the tranversal spin fluctuations appear to
decouple in the infrared}. Although this is theoretically not well understood,
this appears to be implied by
 the fact that the charge- and spin ordering
transitions are decoupled\cite{tran,tranni}. This can be stated more precisely
as follows: as long as the charged domain walls are intact, the spin system
is unfrustrated. Calling the exchange interaction inside the spin domains $J$
and that between spins on opposite sides of the wall $J'$, the (classical)
spin system appears identical for every possible configuration of domain walls
when $J=J'$. Apparently, $J-J'$ is an irrelevant operator. Hence, although
microscopically $J \neq J'$ (and frustration), the spin- and domain wall
dynamics decouple in the approach to the long wave length limit, although the
numerical factors (like the spin wave velocity) have to be adjusted. 

Does this mean that the domain wall dynamics would be completely invisible
in a measurement of the spin dynamical form factor? In fact,
in this way one measures a {\em convolution} of (transversal) spin- and domain
wall dynamics. One can define two distinct equal time correlators ($\vec{M}_i =
(-1)^i \vec{S}_i$ is staggered spin), 
\begin{equation}
S ( R ) = \langle \vec{M}_{i+R} \cdot \vec{M}_i \rangle \sim e^{-R/\xi}
\end{equation}
\begin{equation}
S_T ( R )  =  \langle \vec{M}_{i + R} \left[ \Pi_{l=i+R}^{i} e^{i \pi \hat{n}_l} \right]
\vec{M}_i \rangle \sim e^{-R/\xi_T}
\end{equation}
where the last proportionalities hold when the spin system is disordered. In
Eq. (4), $\hat{n}$ measures the charge associated with a single domain wall
unit cell (e.g., in the nickelates $\hat{n}$ corresponds with the hole number
operator, while in the half-filled cuprate walls $\hat{n}=1$ corresponds with
half a hole). Hence, the charge sector appears in a {\em disorder operator}
form in the spin correlator Eq. (4): regardless of the location of the domain 
walls, the change of sign of the spin order parameter upon crossing the walls
is undone by the term between square brackets. 
It is this correlator which measures the transversal spin
correlations. On the other hand, in the experiment one measures the
correlator Eq.(3), 
which is also sensitive for the disorder induced by the domain
wall fluid. This general wisdom should also apply to unequal time correlations. 
   
With the above observations, it becomes possible to deduce 
some characteristics of the
$\vec{q}$ and $\omega$ dependence of the spin dynamical structure factor.
Let us consider the case that Eq. (4) would show near-long-range order,
while the charge sector is an incommensurate fluid ($\xi_T >> \xi$). To
address the temporal aspects of the dynamics it is convenient to consider
the imaginary part of the {\em local} dynamical spin susceptibility
$\chi'' (\omega) = \int d\vec{q} \chi'' ( \vec{q}, \omega )$: every
time a domain wall passes a particular site, the spin at this site is
flipped and $\chi'' (\omega)$ is dominated by this dissipation source.
It is easy to see that
the domain wall scale $\Gamma$, Eq. (2), should appear as a characteristic
scale in $\chi''$: (a) when $\omega >> \Gamma$ the domain wall system
appears as static and the frequency dependence should be similar to that
of the pure spin system, (b) when $\omega \rightarrow 0$ the spin-flip rate
is non-zero as long as the domain walls are in a fluid state (even when
$\xi_T \rightarrow \infty$) and on hydrodynamic grounds $\chi'' (\omega)
\sim \omega$. In ref.\cite{zhvs} we present a more detailed hydrodynamic
argument showing that in fact $\chi'' (\omega, T) \sim \omega / T$. The
crossover regime between the high- and low frequency regimes should be
governed by the single scale parameter $\Gamma$, and $\chi'' ( \omega, T )
\sim F ( \omega / \Gamma) = F (\omega / (\mu k_B T )$, and this is the 
scaling observed in neutron scattering \cite{aeppli}. Hence, we suggest
that this crossover regime is actually probed in the inelastic neutron
scattering experiments. As an interesting ramification, the prefactor 
$\mu$ in Eq. (2) is also the dimensionless quantum of action of the
incommensurate fluid: the characteristic energy scale for quantum 
fluctuations 
$\hbar \Gamma_Q = ( \hbar c_W / a ) \exp ( -1 / \mu )$\cite{zhvs}.
If, and only if, $\mu$ is of order unity a crossover is expected
at a reasonable temperature from a high temperature classical
fluid to a low temperature quantum regime (like in $^4He$) and this
appears as a consistency requirement in the present context. Experimentally,
$\mu \simeq 1$ \cite{aeppli}.

A little more can be said regarding the spatio-temporal appearance of the
magnetic fluctuations. Assuming the near decoupling of the transversal-
and domain wall dynamics, the spin wave (transversal mode)
spectrum of the static striped phase appears as a sensible zeroth order:
the goldstone modes attach to the incommensurate
wave vectors and the branches should come together at the $(\pi, \pi)$ point,
while gaps $\sim J-J'$ are found at the new Brillioun zone boundaries at
smaller momenta\cite{tranni1}. When the
domain walls are disordered, one expects that the spin waves start to
propagate when their wave length becomes of order of, 
or less than the disordering
length of the domain wall fluid. Longer wavelength spin waves will be 
overdamped. Hence, as a function of frequency a crossover has to occur from
an overdamped response at low frequencies to an underdamped spin wave response
at higher frequencies. The frequency where this occurs is $\omega_c \sim 
c \pi / l_{m.f.}$, where the mean-free path of the spin waves is expected
to be proportional to the domain wall disordering length: $l_{m.f.} \simeq
\alpha \xi_c$, where $\alpha$ is a spin wave-domain wall
scattering cross section ($0 < \alpha < 1$). 
At frequencies much less than $\omega_c$, the width in momentum space 
should be set by $\pi / l_{m.f.}$. As long as the spin-wave cones are
unresolved,
the width of the peaks in $S(\vec{q}, \omega)$ should behave as 
a simple geometric average
of the inverse domain wall disordering length ($1 /(\alpha \xi_c)$)
and the apparent $\omega$ dependent width coming from the 
unresolved 
spin-wave cone ($\kappa_{\omega} = \omega / c$). The domain wall
disordering length is nothing else than the collision length (Eq.1) and
we expect for the frequency- and temperature dependence of the width
of the incommensurate peaks,
\begin{equation}
\kappa (\omega, T) = \sqrt{ { {(k_B T)^2} \over { (\alpha \Sigma d^2)^2 } }
+ { {\omega^2} \over {c^2} } }
\end{equation}
where $c$ is the spin-wave velocity. 

Except for a residual width, this is the behavior
found in the experiment\cite{aeppli1}. 
Interestingly, Aeppli {\em et al} find that temperature
and frequency appear with common prefactors. Together with $\mu \simeq 1$
this implies that $ \alpha c_W \simeq c$: the product of the spin wave-domain
wall cross section and the wall transversal sound velocity is of order of
the spin wave velocity ($c \simeq 0.2$ eV \AA). This observation might hint at
the microscopic mechanism of stripe formation. 

Further experimentation is needed to see if the above physics is the correct
one. Central to our analysis is the assertion that the collective dynamics
in the normal state is in a {\em classical} regime. General arguments 
indicate that the difference between Eqs. (3,4) acquires a different meaning
in the quantum system, and it is tempting to speculate that
the coherent spin mode observed by Mason et al \cite{mason} reflect the onset
of quantum coherence in the domain wall fluid upon entering the superconducting
regime.

Aknowledgements. We acknowledge stimulating discussions with G. Aeppli,
A. V. Balatski, D. Pines, M. Horbach, S. Kivelson, D. H. Lee and J. Tranquada.
JZ acknowledges financial support by the Dutch Academy of Sciences (KNAW).

\end{multicols}

\end{document}